\input phyzzx
\input epsf

\font\bigboldii=cmbx10 scaled\magstep2
\font\smallitii=cmmi7 scaled\magstep2
\def\gev{~\hbox{GeV}}
\def\tev{~\hbox{TeV}}
\def\up#1{^{\left( #1 \right) }}
\def\lesim{\,{\raise-3pt\hbox{$\sim$}}\!\!\!\!\!{\raise2pt\hbox{$<$}}\,}
\def\gesim{\,{\raise-3pt\hbox{$\sim$}}\!\!\!\!\!{\raise2pt\hbox{$>$}}\,}
\def\sm{Standard Model}
\def\expel{\par\vfill\eject}
\def\lowti#1{_{{\rm #1 }}}
\def\lcal{{\cal L}}
\def\ocal{{\cal O}}
\def\su#1{{SU(#1)}}
\def\ui{U(1)}
\newcount\fignum
\def\thecaphyz#1{\global\advance\fignum by 1
\centerline{\hfil\vbox to 1 in{\hsize 5 in \vfil
{$\mskip -52.24mu$\textindent{\smallitii Fig:{\rm\ }
\the\fignum{\rm\ }}
        \global \advance \baselineskip by -10 pt
        \tenpoint\noindent #1 }}\hfil}\global \advance \baselineskip by 10 pt
\bigskip}
\def\wcal{{\cal W}}

\def\thetitle{CP VIOLATION THROUGH EFFECTIVE LAGRANGIANS}
\def\theabstract{A model independent study of non-\sm\ CP-violating
processes is presented with emphasis on the observability of the effects.}

\def\whatjournal{P}

\def\ordernpb#1#2#3{{\bf#1} (#3) #2}
\if P\whatjournal {\global\def\order#1#2#3{\orderprd{#1}{#2}{#3}}}
     \else
                  {\global\def\order#1#2#3{\ordernpb{#1}{#2}{#3}}}
\fi

\def\app#1#2#3{{\it Acta Phys. Pol. {\bf B}}\order{#1}{#2}{#3}}

\def\ijmpa#1#2#3{{\it Int. J. of Mod. Phys. {\bf A}}\order{#1}{#2}{#3}}

\def\npb#1#2#3{{\it Nucl. Phys. {\bf B}}\order{#1}{#2}{#3}}

\def\plb#1#2#3{{\it Phys. Lett. {\bf B}}\order{#1}{#2}{#3}}

\def\prl#1#2#3{{\it Phys. Rev. Lett.\ }\order{#1}{#2}{#3}}

\def\prd#1#2#3{{\it Phys. Rev. {\bf D}}\order{#1}{#2}{#3}}

\def\zphys#1#2#3{{\it Z. Phys. {\bf C}}\order{#1}{#2}{#3}}

\REF\reviews{ For some recent reviews see
G. Valencia, contributed to {\sl TASI 94} (hep-ph/9411441)
R.D. Peccei, presented at {\sl Workshop on Physics and
Detectors at DAPHNE}, Frascati, Italy, Apr 4-7, 1995. (hep-ph/9508389).}
\REF\models{For example,
B. Grzadkowski and J.F. Gunion, \plb{350}{218}{1995}
A. Acker \etal, \prd{48}{5006}{93}.
F. del \'Aguila \etal, \plb{129}{77}{1983}.}
\REF\phen{See, for example, Brandenburg \etal, \zphys{51}{225}{1991}.}
\REF\leff{
H. Georgi, {\it Annual review of nuclear and particle science} {\bf43}, 209
(1994).
J. Wudka, \ijmpa{9}{2301}{1994}.}
\REF\veltiruv{M. Veltman, \app{12}{437}{1981}.}
\REF\bw{ W. Buchmuller and. Wyler, \npb{268}{621}{1986}.}
\REF\frere{ J.M. Fr\`ere \etal, \plb{292}{348}{1992}.}
\REF\perez{M.A. P\'erez \etal, \prd{52}{494}{1995}.}
\REF\estimates{H. Georgi, \plb{298}{187}{1993}.}
\REF\wudka{J. Wudka, Ref \leff}
\REF\aew{ C. Arzt \etal, \npb{433}{41}{1995}.}
\REF\eeww{ K. Hagiwara \etal, \npb{282}{253}{1987}.}
\REF\meaningofbounds{A. De R\'ujula \etal, \npb{384}{3-58}{1992},
C. Arzt \etal, \prd{49}{1370}{1994}, J. Wudka, Ref. \leff.}
\REF\eom{C. Arzt, \plb{342}{189}{1995}.}
\REF\dec{ T. Appelquist and J. Carazzone (Harvard U.).\prd{11}{2856}{1975}.}
\REF\buch{ W. Buchmuller \etal, \plb{197}{379}{1987}.}
\REF\pdg{Part. data group, \prd{50}{1175}{1994}.}
\REF\naturality{ G. 't Hooft, lecture given at {\sl Cargese Summer
Inst.}, Cargese, France, Aug 26 - Sep 8, 1979.}
\REF\kkdif{G. Beall \etal, \prl{48}{848}{1982},
G.C. Branco \etal, \npb{221}{317}{1983}.}
\REF\nedm{J.M. Fr\`ere \etal, \plb{251}{443}{1990}.}
\REF\valencia{For a review see, for example, Valencia, Ref. \reviews}
\REF\nacht{M. Diehl and O. Nachtmann, \zphys{62}{397}{1994}.}
\REF\gw{B.Grzadkowski and J. Wudka, \plb{364}{49}{1995}.}
\REF\ewnew{M.B.Einhorn and J.Wudka in preparation}
%

%\noblackbox
%\nopagenumbers

\baselineskip 18 pt

\advance \hsize by 2.1 cm
\advance \hoffset by -0.2 cm
\advance \vsize by 4 cm
\advance \voffset by -2.6 cm

\def\refmarkforthis#1{\attach{\scriptscriptstyle   #1 ) }}

\null
\vskip 3.5 cm

\centerline{{\bigboldii \thetitle}}
\vskip 36pt
\centerline{{\it Jos\'e Wudka}}
\centerline{{Physics Dept., U.C. Riverside. Riverside, CA 92521-0413,
U.S.A.}}

\vskip 151 pt

\centerline{\vbox{\theabstract}}

\expel
\vskip -16 pt\chapter{Introduction}\vskip -8 pt

The origin of CP violation is one of the important unanswered questions
in particle physics despite the enormous attention the subject has
received~\refmarkforthis{\reviews}. The question which I will address in this
lecture is what kind of CP violating effects can we expect from
non-\sm\ physics, what type of new physics can generate such effects,
and whether they can be observed at present and near-future colliders.

There have been many studies of CP violation for specific
models~\refmarkforthis{\models}. There have also been some attempts to obtain
model-independent statements concerning CP violation~\refmarkforthis{\phen}.
The formalism which I will use is based on a gauge-invariant effective
Lagrangian approach~\refmarkforthis{\leff} which provides not only a consistent
framework for this study but also provides estimates of the magnitude of
the effects under consideration.

\vskip -16 pt\chapter{Effective Lagrangians}\vskip -8 pt

Consider a theory containing a set of light fields $ \phi $
and a set of heavy fields $ \Phi $ described by
the action $ S [ \phi ; \Phi ]  $. Suppose also that we cannot directly
observe the heavy physics which becomes manifest at a scale $ \Lambda $.
In this case heavy physics can be observed using only {\it virtual} heavy
effects which are described by the effective action $ S \lowti{eff} $ defined
by
$ \exp \left( i S\lowti{ eff } [ \phi ] \right) = \int [ d\Phi ] e^{ i
S } $. Expanding $ S\lowti{eff} $ in powers of $ 1 /
\Lambda $~\refmarkforthis{\leff},
$ S \lowti{eff} = \int \lcal\lowti{eff} $ defines the
defines the effective Lagrangian $$ \lcal\lowti{eff} = \sum_n { \alpha_n
\Lambda^n } \ocal_n . \eqn\eq$$ in terms of a series of {\it local}
operators $ \ocal_n $.
If all terms of dimension $ \le 4 $ have a
local symmetry then either that symmetry is preserved by {\it all} the
operators, or else the renormalization group will generate terms of
dimension $ \le 4$ which break this symmetry~\refmarkforthis{\veltiruv}.
If we assume that the terms in $ \lcal\lowti{eff} $ of dimension $ \le 4
$ correspond to the \sm, it follows that we must assume that all $ \ocal_n $
are $ \su3 \times \su2 \times \ui $ invariant.~\foot{Similar statements
do not apply to global symmetries.}
To complete this parameterization one requires the list of light fields;
and currently we have two possibilities depending on the presence or
absence of light scalars. In this talk I will assume that the light spectrum
coincides
with the one in the \sm (including a Higgs
doublet)~\refmarkforthis{\bw}.~\foot{If there
are no light scalar excitations a chiral Lagrangian description of the
theory is appropriate~\refmarkforthis{\reviews}}

Any kind of new physics can be parameterized by the $ \alpha_n $ which
summarize all the virtual heavy-physics effects. Any
experiment which do not probe the new physics directly can glean
information about the new interactions only by measuring these
coefficients.

It is, of course, possible to assume that the effective operators of
dimension $ > 4 $ satisfy a larger local symmetry than the one of the
\sm~\refmarkforthis{\frere}, it is also possible to consider more complicates
light scalar sectors\refmarkforthis{\perez}. I will not consider these cases
for simplicity.

\vskip -16 pt \chapter{Phenomenological estimates}\vskip -8 pt

The coefficients $ \alpha_n $ can be constrained by requiring consistency
of the theory\refmarkforthis{\estimates}. For the case under
consideration (where there are light scalars), I will assume that the
underlying theory is weakly coupled~\refmarkforthis{\wudka}. Then the relevant
property of a given operator is whether it can be generated at tree
level by the heavy physics~\refmarkforthis{\aew}: all tree-level-generated
operators have
coefficients equal to some product of the coupling constants,
loop-generated operators have additional suppression
factors~\foot{When there are many ($\sim 150$) loop graphs
which add coherently to cancel this loop suppression factor the low
energy spectrum is also modified since the theory becomes strongly
coupled: the one loop and the tree level graphs are of the same order.
See~\refmarkforthis{\wudka} for more details.}. $
\sim 1/ ( 4 \pi )^2 $. I will also assume that gauge fields are
universally coupled. These considerations lead top the estimates
presented in figure 1; such estimates are also verified in explicit
calculations. With this estimates $ \Lambda $ is identified as a
physical mass scale (eg. the mass of a heavy particle).

\setbox2=\vbox to 50 pt {\epsfysize=3.5 truein\epsfbox[120 -70 732 722]{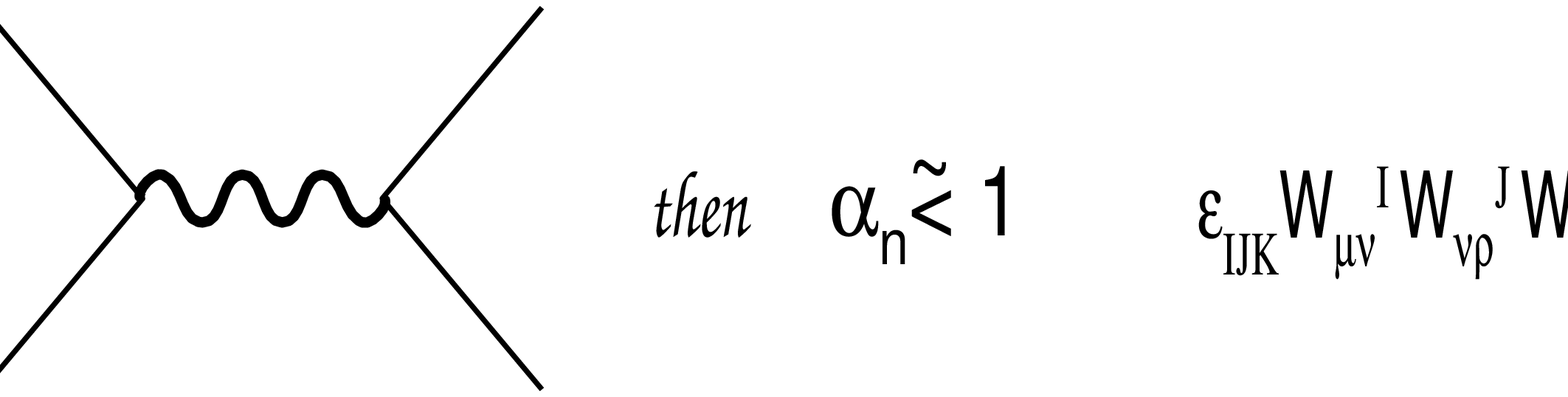}}

\centerline{\box2}

\thecaphyz{Examples of operators generated only via loops (right) and
operators that can be generated via tree level processes (left).}

Using the above estimates one can determine whether a given experimental
bound does constrain a theory or not. For example consider the
$WW\gamma$ interaction~\refmarkforthis{\eeww} (in unitary gauge),
$$ \lcal_{ WWV} =  { i e \lambda
\over M_w^2 } \wcal^+_{\alpha \mu } \wcal^-_{ \nu\mu } F^{\lambda \nu },
\quad \wcal^\pm_{ \mu \nu }  =
\partial_\mu W^\pm_\nu - \partial_\nu W^\pm_\mu
\eqn\eq $$ where $ W^\pm $ denote the
$W$-boson fields, and $F$ the usual photon field strength
which is generated by the
operator $\ocal_W = \epsilon_{ I J K } W^I_{ \mu \nu }
W^J_{ \nu \lambda } W^K_{\lambda \mu } ,$ so that $$
\lambda_V = { 3 g v^2 \over 2 \Lambda^2 } \alpha_W \eqn\eq $$ where $v$ denotes
the \sm\ vacuum expectation value, $g$ the $ \su2_L$ gauge coupling
constant and $ \alpha_W $ the coefficient of $ \ocal_W $ in the
effective Lagrangian. Since $ \ocal_W $ is only
generated via loops~\refmarkforthis{\aew} in the underlying theory we expect $
\alpha_W
\sim g^3/(16 \pi^2 ) $. A given bound on $ \lambda $
can now be translated into a constraint on $ \Lambda $, for example
$$ | \lambda_V | < 0.1 \quad \Rightarrow
\quad \Lambda > \hbox{few}\times 10 \gev , \eqn\eq $$ so that one cannot
claim consider this a high precision
measurement~\refmarkforthis{\meaningofbounds}

Within specific models it is possible to find $ \alpha_n $
enhanced or suppressions (perhaps due to unknown symmetries)
with respect to the above estimates. Still
one cannot assume with impunity  for $\alpha_n $ are enhanced
by many orders of magnitude: such enormous
discrepancies would have observable consequences in other processes and
would have been detected
Note that the same statement can be made about the gluon operator
studied in~\refmarkforthis{\phen}.

\vskip -16 pt \chapter{CP violating operators}\vskip -8 pt

I will now consider the operators which do not respect CP for the case
where the underlying theory is weakly coupled and decoupling. The light
excitations will be again those of the \sm\ with one scalar doublet.
In this case all operators of dimension $ \le6 $ are
known~\refmarkforthis{\bw}.
{}From the above arguments I only need those operators that
can be generated by tree-level graphs~\foot{I will also use the
equations of motion to eliminate operators that are indistinguishable at
the level of the $S$ matrix~\refmarkforthis{\eom}}.

\setbox2=\vbox to 100 pt{\epsfysize=4 truein\epsfbox[0 -80 612 672]{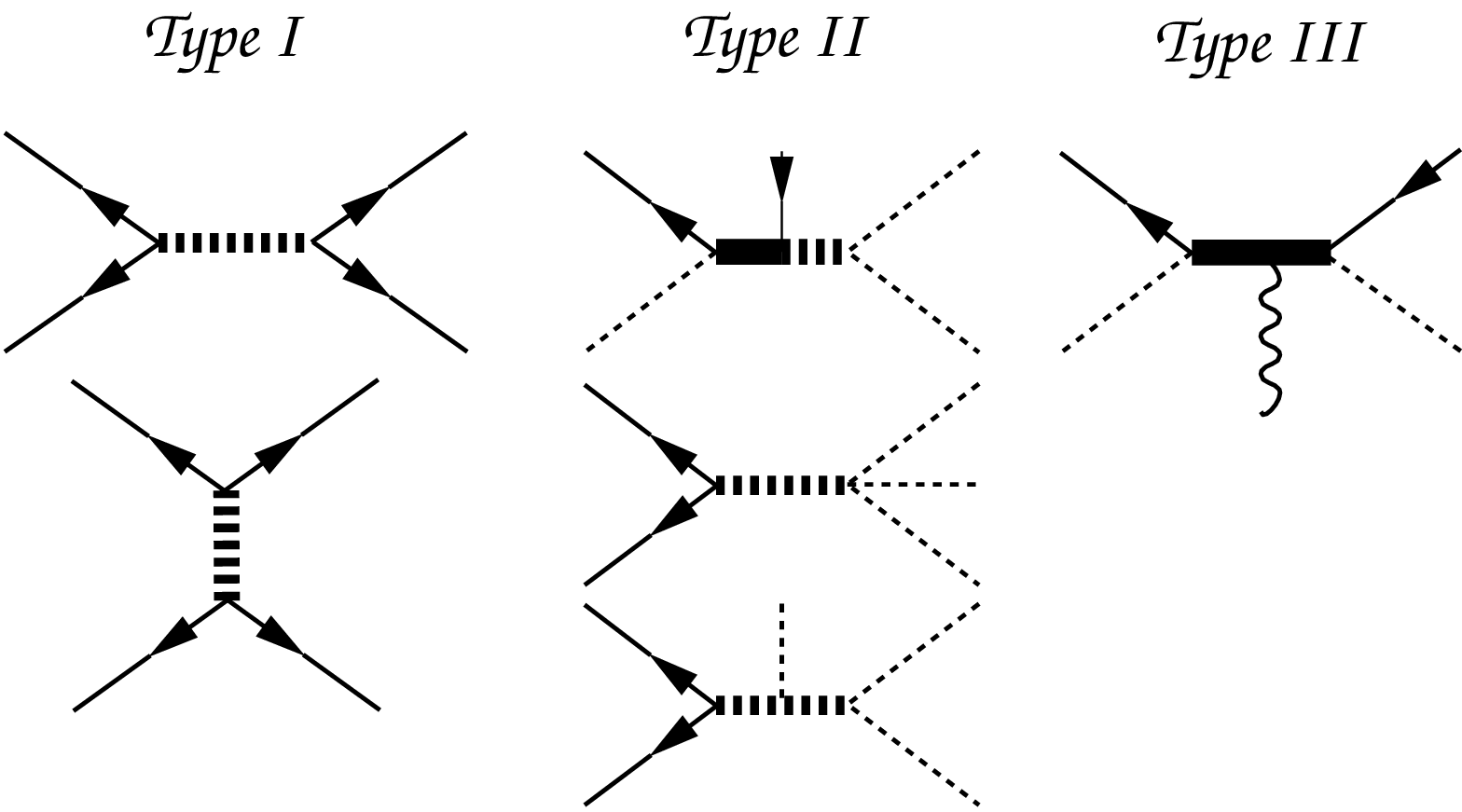}}

\centerline{\hbox{\box2}}

\thecaphyz{Decoupling physics responsible for tree-level generated,
CP-violating operators. heavy solid ()dashed) lines denote heavy fermions
(scalars).}

The number of operators is surprisingly small, they fall into three
categories:
$$ \eqalign{
& \left( \bar\psi_L\up1 \psi_R\up2 \right)
\left( \bar\psi_{L,R}\up3 \psi_{R,L}\up4 \right) - \hbox{H.c.} \ : \ \hbox{
Type\ I} \cr
& \left( \bar \psi\up1_L \psi\up2_R \phi \right) \left( \phi^\dagger
\phi \right) - \hbox{H.c.} \ : \ \hbox{ Type\ II}  \cr
& \left( \phi^T \epsilon D_\mu \phi \right) \left( \bar\psi\up1_{ L , R }
\gamma^\mu  \psi\up2_{ L , R } \right)  - \hbox{H.c.}  \ : \
\hbox{ Type\ III} \cr } \eqn\CPops $$
where $ \psi_L $ denote left-handed fermion doublets, $ \psi_R $
right-handed fermion singlets, $ \phi $ represents the scalar doublet, $
D_\mu $ the covariant derivative and $ \epsilon = i \sigma_2 $ is the $
2 \times 2 $ antisymmetric matrix of unit determinant. In \CPops\ the
fields are restricted by the condition that the total hypercharge
should be zero in order to preserve gauge invariance.

The various types of heavy physics responsible for the operators of
types I, II and III are given in figure 2 above (since all
heavy physics effects vanish as $ \Lambda \rightarrow \infty $ this
type of new physics is labeled ``decoupling''~\refmarkforthis{\dec}).
Such operators appear in the effective Lagrangian with unknown
coefficients (bounded by the requirement that they are $ \lesim 1 $).
There are, however, some experimental bounds on the $\alpha_n$.

$\star$
\undertext{{\sl Operators of type I.}}
When the operators involve first generation fermions only $
\Lambda \gesim 10 \tev  $ ($\alpha \sim 1 $)
from $ \pi , K \rightarrow e \nu
$~\refmarkforthis{\buch} and from the
electron and neutron dipole moments~\refmarkforthis{\pdg}.
When the operators involve second and third generation
fermions the bounds are weak ($ \Lambda \gesim 10 \gev $).
These operators also contribute
at one loop to the $ \theta $
parameter. If we require the theory to be natural~\refmarkforthis{\naturality}
(at least where $ \theta $ is concerned) then $ \Lambda \gg 10 \tev $.

$\star$
\undertext{{\sl Operators of type II.}} For first generation fermions
$ \Lambda \gesim \hbox{(few)} \times 100 \gev $ ($ \alpha \sim 1 $)
from the electric and magnetic dipole moments of the leptons
and neutron, using a Higgs mass~\foot{These bounds are
generated by loop graphs involving these operators and therefore
depend on the Higgs mass.} $ \sim 100 \gev $. When these operators
involve the second and third generation fermions the bounds are very
weak. These operators also modify to the $ \theta $ parameter at
tree-level for natural~\refmarkforthis{\naturality} theories
$ \Lambda > 10^4 \tev $.

$\star$
\undertext{{\sl Operators of type III.}}
When involving first-generation fermions only bounds can be obtained
using the $W$ lifetime and branching ratios, the $ K_L - K_S $ mass
difference~\refmarkforthis{\kkdif} which leads to $
\Lambda \gesim 500 \gev $; a bound using the neutron edm~\refmarkforthis{\nedm}
is polluted by the presence of unknown angles.
 When these operators
involve the second and third generation fermions the bounds are very
weak. These are the least constrained operators, processes affected by
these operators may then be particularly sensitive to heavy CP violating
effects.  This type of
operators are generated {\it only} by a heavy fermion isodoublet
of non-zero hypercharge.
It is interesting to note that this type of heavy fermions would suppress
the $ Z \rightarrow b \bar b $ branching ratio .

\vskip -16 pt \chapter{Conclusion}\vskip -8 pt

For the case of decoupling heavy physics the best windows into new
types of CP violation is through those observables sensitive to three
types of operators: 4-fermion operators, operators modifying the
fermion-Higgs couplings and operators modifying the $ W t b $ and $ W t
b H $ couplings. If the underlying theory is also assumed to satisfy the
usual naturality criteria~\refmarkforthis{\naturality} then only the operators
modifying the $W$ couplings could be generated by physics light enough
to be of interest in near-future collider experiments.

Having a CP violating terms in the Lagrangian is, unfortunately,
not enough. In order to probe the CP violating effects
one must construct observables containing the corresponding coefficients.
Such observables are either proportional to the interference of some CP-even
phase with the CP violating phase~\refmarkforthis{\valencia}, or are obtained
by averaging a CP-violating quantity~\refmarkforthis{\nacht}. In both cases the
effects are considerably suppressed

Thus, even when $ \Lambda $ is sufficiently small for the effects of the
heavy physics to be observable at a given collider, the CP-violating
effects would be very hard to observe: the CP violating couplings (for
the case of the fermion-$W$ interactions) are $ \sim g ( v / \Lambda )^2
$;if we take the LEP bounds of $ \Lambda \gesim 2 \tev
$~\refmarkforthis{\gw} this is reduced to $ \lesim g/64 $.

I would like to conclude by noting that a similar investigation can be
done in the case where there are no light scalars by using  a chiral
effective Lagrangian~\refmarkforthis{\ewnew}. Finally one might wonder what
would happen if
the underlying theory is both decoupling and strongly coupled. In this
case (which I completely ignored) it is difficult to maintain the Higgs
mass significantly below the cutoff requiring fine-tuning. The
alternative is to modify the low-energy spectrum. I will consider all
there possibilities  in a forthcoming publication.

\ack
I would like to thank M. Einhorn and J.M Fr\`ere for many useful comments.
This work was  supported in part through funds provided by the Department
of Energy

\tenpoint
\baselineskip 15 pt
\refout

\bye